\documentstyle[12pt,epsfig]{article}
\def\beq{\begin{eqnarray}}
\def\eeq{\end{eqnarray}}

\begin{document}
\renewcommand{\thefootnote}{\fnsymbol{footnote}}
\begin{titlepage}
\begin{flushright}
\end{flushright}
\vskip0.5cm
\begin{center}
{\LARGE\bf
Renormalizaion group equations and infrared quasi fixed point behaviors of 
non-universal soft terms in MSSM} 
\vspace*{0.5cm}

       {\bf Chao-Shang HUANG}$^{a,c}$\footnote{E-mail : csh@itp.ac.cn},
        {\bf LIAO Wei }$^{a,c}$\footnote{E-mail: liaow@itp.ac.cn},
{\bf Qi-Shu YAN}$^b$\footnote{E-mail : qsyan@mail.tsinghua.edu.cn},
  and
       {\bf Shou-Hua ZHU}$^a$\footnote{E-mail: huald@itp.ac.cn}
        \vspace{0.5cm}

        $^a$Institute of Theoretical Physics,
         Academia Sinica, 100080 Beijing, China \\
        $^b$ Physics Department of Tsinghua University, 100080 Beijing, China \\

        $^c$ The Abdus Salam International Centre for Theoretical Physics, 

         P.O.Box 586, 34014 Trieste, Italy \\
\bigskip
\end{center}

{\large\bf}
\centerline{ }
\centerline{ }
\bigskip
\normalsize

\begin{abstract}
  The renormalization group equations(RGEs) of non-universal soft 
supersymmetric breaking terms with CP violating phases are analyzed
in this paper. We obtain the analytic solutions of RGEs by directly
solving the RGEs themselves. Compared with the method of spurion
expansion our approach proves to be simple and succinct, and easy
to extend to the case of complex parameters. With the analytical
forms of the solutions we obtained the infrared quasi fixed point 
behavior of soft terms are analyzed and it turns out to support the
notion in scenarios with CP violating phases.
\end{abstract}
\bigskip

\end{titlepage}

\section{Introduction}
~~~~As an excellent candidate to embed the standard model(SM) in a more
fundamental theory, the minimal supersymmetric standard model(MSSM)
has many good features. It ingeniously tackles the abominable gauge
hierarchy problem from which ordinary unification theories suffer
\cite{soh}, provides a mechanism that supergravity models all
share which breaks the electroweak(EW) symmetry dynamically via
radiative corrections, naturally provides the lightest supersymmetric
particle(LSP) as a candidate of the dark matter in astrophysics and
cosmology\cite{jkg}, etc. In spite of these theoretical virtues,
however, the MSSM suffers from great uncertainty which arises from
the large number of free parameters describing the soft supersymmetry
(SUSY) breaking. One may greatly reduce the parameter space and make
the theory much more predictive via adopting the universality
conditions at high energy scale, e.g., one may choose the minimal
supergravity model(MSUGRA). Furthermore, with the aid of the concept of
infrared quasi fixed point(IRQFP), further reduction of parameter space
has been observed in many works\cite{hill,lot,cw,ye}. It reveals the screening
effect for soft breaking parameters implied entirely by the renormalization
group equations(RGEs) and the large top Yukawa coupling, and with this 
one may rely on two\cite{cw} or even one\cite{ye} free soft SUSY breaking
terms, i.e., $M_0$ and $M_{1/2}$ or only $M_{1/2}$. In these cases one
can work in models of highly predictive power leaving some of the
universality boundary assumptions. This IRQFP analysis was extended
to the large tan$\beta$ scenario\cite{jk} in which the whole set of
Yukawa coupling of the third generation has to be considered. The result
shows that the theory also exhibit the IRQFP for large tan$\beta$ and
allows one to predict SUSY mass spectra as functions only of $M_{1/2}$,
 but the analysis turns out to be much more complicated.

While hindered in this situation, a crucial observation was made in
Ref. ~\cite{kaz} which showed the intrinsic connection between the
solutions for the RGEs of soft terms and those of the gauge coupling
constants and Yukawa couplings. It tells that one can get the solutions of RGEs for
the soft SUSY breaking terms via substituting the modified expressions for
gauge and Yukawa couplings in the solutions of the RGEs for gauge and Yukawa couplings
\cite{kaz,av,ck}.
\beq
\alpha_i &\Rightarrow& \tilde{\alpha}_i=\alpha_i(1+M_i \eta+\bar M_i
\bar{\eta}+2M_i\bar M_i \eta \bar{\eta}),\ \ \ \eta=\theta^2, \ \ \
\bar{\eta}=\bar{\theta}^2, \label{g} \nonumber \\
Y_k &\Rightarrow & \tilde{Y}_k= Y_k(1-A_k\eta -\bar{A}_k\bar{\eta}
+A_k\bar{A}_k\eta\bar{\eta} +\Sigma_k\eta\bar{\eta}) \ , \label{Y} \nonumber
\eeq
and expanding over the Grassmannian parameters $\theta$ and $\bar{\theta}$,
which is exactly the spurion field description of soft SUSY breaking
\cite{gg}. The technique was immediately used to probe the implication of
RGEs for the soft SUSY breaking terms in large tan$\beta$ scenario, but
one faced with the problem of finding good analytical forms for 
Yukawa couplings. This problem was solved in Ref.~\cite{am} in which 
an integral form for Yukawa coupling evolution was given which was
shown to be a convergent scheme for numerical calculations and a
convenient and concise form for qualitative analysis. In view of this
solution, the general analytical forms for the solutions of the soft
SUSY breaking terms were obtained in this complicated case via the 
spurion field expansion and the IRQFP feature was analyzed for 
non-universal boundaries of soft terms\cite{km}.

Although the solution for soft SUSY breaking terms look concise and
simple, the procedure described above, especially the actual 
calculation of spurion expansion is still complicated due to the
a little bit complicated iterative solutions for Yukawa couplings
\cite{am}. So in Ref.~\cite{km} the author restrict their attention to
real trilinear couplings and real gaugino masses. However, in a lot of works on
superstring theory and M-theory phenomenology non-universal complex soft terms 
appear naturally~\cite{mp}, which provide new sources of CP violation. And the
SUSY CP violating phases may not be constrained to be small due to the cancellation 
mechanism of SUSY contributions to electron and neutron electric dipole moments
(EDME and EDMN), and one has extensively discussed the phenomenological 
implications of large phases of $\mu$, gaugino masses and
trilinear terms while satisfying the EDME and EDMN constraints~\cite{ipb}. 
Therefore, it is important to extend the study of non-universality of soft terms
to the complex parameter case.
In this paper we will show that the convenient iterative forms\cite{am,km}of the 
solutions of RGEs can be obtained very succinctly in the approach of directly solving
RGEs and extend to the complex non-universal soft terms. Our results show that the real
and imaginary parts of trilinear couplings have infrared quasi fixed points respectively
if the initial values of Yukawa couplings are large enough ( at least $Y^0_k > \alpha^0$).

The paper is organized as follows. In section two we will
show briefly how one can directly solve the RGEs and arrive at the 
forms in Ref.~\cite{km}, and extend the analysis to more general case, complex parameter
case. In section three we will analyze the IRQFP in the case with SUSY CP violating phases
and discuss the phenomenological implications. Finally, conclusions are drawn in section four.

\section{Solution of the RGEs }
~~~~The one loop RGEs for gauge and Yukawa coupling and soft breaking terms can be
written as
\beq
\dot{\alpha}_i &=& -b_i\alpha_i^2, \\ 
\dot{Y}_k &=&
Y_k(\sum_{i}c_{ki}\alpha_i - \sum_{l}a_{kl}Y_l),\\
\dot{M}_i &=& -b_i\alpha_i M_i,\\
\dot{A}_k &=&
-\sum_{i}c_{ki}\alpha_i M_i-\sum_{l}a_{kl} Y_l A_l,\\
\dot{\Sigma}_k &=&
2 \sum_{i} c_{ki}\alpha_i |M_i|^2-\sum_{l}a_{kl} Y_l(\Sigma_l+|A_l|^2),
\eeq
where  $\alpha_i= g_i^2/16\pi^2$, i=1,2,3, $Y_l=y_l^2/16\pi^2$, l=t,b,$\tau, \cdot \equiv d/dt, 
\ t= \log M_{GUT}^2/Q^2$ and
\begin{eqnarray*}
\Sigma_t &=& \tilde{m}_{Q3}^2 + \tilde{m}_{U3}^2  + m^2_{H2}, \\
\Sigma_b &=& \tilde{m}_{Q3}^2 + \tilde{m}_{D3}^2  + m^2_{H1}, \\
\Sigma_\tau &=& \tilde{m}_{L3}^2 + \tilde{m}_{E3}^2 + m^2_{H1}, \\ 
b_i&=&\{33/5,1,-3 \}, \\ c_{ti}&=& \{13/15,3,16/3 \}, \ \
c_{bi}=\{7/15,3,1 6/3 \},\ \  c_{\tau i}=\{9/5,3,0 \}, \\
a_{tl}&=&\{6,1,0 \},\ \ a_{bl}=\{1,6,1 \},\ \ a_{\tau l}=\{0,3,4 \}.
\end{eqnarray*}

The solutions for $\alpha_i$'s and $M_i$'s can be easily obtained to be:
\beq
\alpha_i=\frac{\alpha_i^0}{1+b_i \alpha_i^0 t}, \ \
M_i=\frac{M_i^0}{1+b_i \alpha_i^0 t}.
\eeq
For $Y_k$'s, solutions turn out to be\cite{am,km}
\beq
Y_k & =&  \frac{Y_k^0u_k}{1+a_{kk}Y_k^0\int_0^t u_k}
\eeq
where $u_k$'s are defined iteratively as
\beq
u_k &=& E_k \prod_{l\ne k}(1+a_{ll}Y_l^0 \int_0^tu_l)^{-a_{kl}/a_{ll}},
\eeq 
with 
\beq
E_k= \prod_{i=1}^3(1+b_i\alpha_i^0t)^{c_{ki}/b_i}.
\eeq

To find the analytic solutions of the renormalization group equations for $A_k$'s and 
$\Sigma_k$'s, we rewrite them as the form of the
Bernoulli equation. Consider the trilinear coupling $A_k$ first.
 We can rewrite Eq. (4) to be
\beq
\frac{d A_k}{dt}= -\frac{de_k}{dt}-a_{kk}Y_k A_k,
\eeq
with $e_k$ defined by
\beq
\frac{de_k}{dt} &=& \sum_ic_{ki}\alpha_iM_i+\sum_{l\ne k}a_{kl} Y_l A_l, 
\nonumber \\
e_k^0 &=& e_k(t)|_{t=0}=0.
\eeq 
Eq. (10) can then be solved in the standard way. The solution is
\beq
A_k = -e_k+ \frac{A_k^0+ a_{kk} Y_k^0 \int_0^t u_k e_k}
{1+a_{kk} Y_k^0 \int_0^t u_k}
\eeq
where $A_k^0=A_k|_{t=0}$. In order to solve Eq. (11), we rewrite Eq. (10) as
\beq
\frac{d(A_k+e_k)}{dt}= -a_{kk} Y_k A_k
\eeq
which means that
\beq
\int_0^t Y_k A_k = -\frac{1}{a_{kk}}(A_k+e_k-A_k^0)
=Y_k^0 \frac{A_k^0 \int_0^t u_k-\int_0^t u_k e_k}{1+a_{kk} Y_k^0 \int_0^t u_k}.
\eeq
Integrating Eq. (11) over t and inserting Eq. (14) in it, we find the iterative integral
equations for $e_k$'s
\beq
e_k=t \sum_i c_{ki} \alpha_i M_i^0+ \sum_{l\ne k}a_{kl}
\frac{A_l^0 \int_0^t u_l-\int_0^t u_l e_l}{1/Y_l^0+ a_{ll} \int_0^t u_l}.
\eeq

For $\Sigma_k$'s, the situation is similar but a little more complicated.
From Eq. (13) we can write
\beq
\frac{d|A_k+e_k|^2}{dt}=-a_{kk} Y_k A_k (A_k+e_k)^*-a_{kk} Y_k A_k^* (A_k+e_k),
\nonumber
\eeq
or
\beq
\frac{d(|A_k|^2+A_k^* e_k+A_k e_k^*)}{dt}=-\frac{d|e_k|^2}{dt}-a_{kk} Y_k |A_k|^2
-a_{kk} Y_k (|A_k|^2+A_k^* e_k+A_k e_k^*). 
\eeq
Define 
\beq
\tilde{\Sigma}_k = \Sigma_k -|A_k|^2-A_k^* e_k-A_k e_k^*,
\eeq
then we find from Eqs. (5) and (16) that
\beq
\frac{d\tilde{\Sigma}_k}{dt}=\frac{d\xi_k}{dt}-a_{kk} Y_k \tilde{\Sigma}_k
\eeq
which is the same as Eq. (10) with $A_k$ and $e_k$ substituted by $\tilde {\Sigma}_k$
and $\xi_k$ respectively. In Eq. (18) $\xi_k$'s are defined by
\beq
\frac{d\xi_k}{dt} &=&2 \sum_i c_{ki} \alpha_i |M_i|^2- \sum_{l\ne k}
a_{kl} Y_l (\Sigma_l+|A_l|^2)+\frac{d|e_k|^2}{dt}, \nonumber \\
\xi_k^0 &=& \xi_k(t)|_{t=0}=0.
\eeq
Then from Eqs. (17) and (18) we get
\beq
\Sigma_k=\xi_k+|A_k|^2+A_k^* e_k+A_k e_k^*-\frac{|A_k^0|^2-\Sigma_k^0
+a_{kk} Y_k^0 \int_0^t u_k \xi_k}{1+a_{kk} Y_k^0 \int_0^t u_k}.
\eeq
The iterative equations for $\xi_k$'s are derived in a way similar to that for $e_k$
by noting that Eq. (18) can be rewritten as
\beq
\frac{d\Sigma_k}{dt}=\frac{d\xi_k}{dt}-\frac{d|e_k|^2}{dt}-a_{kk}
 Y_k (\Sigma_k+|A_k|^2), \nonumber
\eeq
which leads to
\beq
\int_0^t Y_k (\Sigma_k+|A_k|^2)=-\frac{1}{a_{kk}}(\Sigma_k-\Sigma_k^0
-\xi_k+|e_k|^2). \nonumber
\eeq
From the above equation and Eq. (19), we find
\beq
&& \xi_k=t^2 |\sum_i c_{ki} \alpha_i M_i^0|^2+2 t \sum_i c_{ki} \alpha_i
|M_i^0|^2-t^2\sum_i c_{ki} b_i \alpha_i^2 |M_i^0|^2  \nonumber \\
&& + t\sum_i c_{ki} \alpha_i M_i^0 \sum_{l\ne k} a_{kl} \frac{A_l^{0*}\int_0^t 
u_l-\int_0^t u_l e_l^*}{1/Y_l^0+a_{ll}\int_0^t u_l}+ t\sum_i c_{ki} \alpha_i
M_i^{0*} \sum_{l\ne k} a_{kl} \frac{A_l^0\int_0^t u_l-\int_0^t u_l e_l}
{1/Y_l^0+a_{ll}\int_0^t u_l} \nonumber \\
&& +\bigg|\sum_{l\ne k} \frac{A_l^0 \int_0^t u_l- \int_0^t u_l e_l}{1/Y_l^0
+a_{ll} \int_0^t u_l}\bigg|^2+\sum_{l \ne k}a_{kl}a_{ll} \bigg| \frac{
A_l^0 \int_0^t u_l- \int_0^t u_l e_l}{1/Y_l^0+a_{ll} \int_0^t u_l}
\bigg|^2  \nonumber \\
&& -\sum_{l\ne k} a_{kl} \frac{(|A_l^0|^2+\Sigma_l^0)\int_0^tu_l-A_l^0
\int_0^t u_l e_l^*-A_l^{0*}\int_0^t u_l e_l+\int_0^t u_l \xi_l}
{1/Y_l^0+a_{ll}\int_0^t u_l}.
\eeq
The Eqs. (12), (15), (20) and (21) are our main results. When $M_i$ and $A_l$
are real they reduce to the results given in Ref.~\cite{km}, as it should be.
The Eqs. (12) and (15) are the same as those in the case of real $A_l$'s and $M_i$'s 
so that it follows that $A_l$ is independent of its
initial value when the initial values of all three Yukawa couplings are large enough 
($Y^0_l$ larger than $\alpha^0$ ). That is, the real and imagine parts of $A_l$ divided by $M_3$
have the IFQFPs, respectively. By inspecting Eqs. (20) and (21), one is led to 
that $\frac{\Sigma_k}{|M_3|^2}$ possesses an IRQFP if the initial values of all three 
Yukawa couplings are large enough. In conclusion Yukawa couplings $Y_l$ and soft terms $A_l$
and $\Sigma_l$ have IRQFP behaviors in the complex non-universal $A_l$ and $M_i$ case when
the initial values of Yukawa couplings, $Y^0_l$'s, are large enough.

It is worth to note that having the analytic solutions of RGEs, it is easy to
find the RG invariants. For example, if we define
\beq
E^\prime_k &=& \prod_{i=1}^3 \alpha_i^{c_{ki}/b_i}, \nonumber \\
u^\prime_k &=& E^{\prime 0}_k u_k = u^{\prime 0}_k u_k \nonumber, \nonumber
\eeq
with $E^{\prime 0}_k=E^\prime_k(t=0)$ and $u^{\prime 0}_k=u^\prime_k(t=0)$,
then from Eqs. (7), (12) and (20) we obtain the RG invariants
\beq
F_{1k} &=& \frac{u'_k}{Y_k}- a_{kk} \int_0^t u'_k, \nonumber \\
F_{2k} &=& \bigg( A_k+ e_k- a_{kk} \frac{Y_k}{u'_k }\int_0^t 
u'_k e_k \bigg) \bigg( 1- a_{kk} \frac{Y_k}{u'_k} \int_0^t
u'_k \bigg)^{-1}, \nonumber \\
F_{3k} &=& \bigg(\Sigma_k- \xi_k- |A_k|^2- A^*_k e_k- A_k e^*_k+
A_{kk} \frac{Y_k}{u'_k} \int_0^t u'_k e_k \bigg) \bigg(1- a_{kk}
\frac{Y_k}{u'_k} \int_0^t u'_k \bigg)^{-1}, \nonumber
\eeq
where $k=t,b,\tau$, which extend the results in Ref. ~\cite{ky} to the case 
with multi Yukawa couplings.

\section{Numerical Study of Dependences of Soft Terms on Their Initial Values }
In order to see the dependences of soft terms on their initial values
one can write

\beq
A_k(t) &=& \sum_l c^k_l(t) A_l^0 + \sum_i d^k_i(t) M^0_i, \\
\Sigma_k(t) &=& \sum_{\alpha}f^k_{\alpha}(t) (m_{\alpha}^0)^2 + \sum_{ll^{\prime}}^k
g^k_{ll^\prime}(t) A^{0*}_l A^0_{l^{\prime}}
+ \sum_{i,l}h^k_{il}(t)(M^0_i A_l^{0*}+ c.c.) +\sum_{i,j}I_{ij}^k(t)M^0_iM_j^{0*},
\label{ini}
\eeq
where the superscript 0 of $A_l^0$, $M_i^0$ etc. means that they are the initial 
values, i.e., the values at unification scale, the asterisk, *, means the complex
conjugate,  $g^k_{ll^{\prime}}=g^k_{l{^\prime}l}$ and $I^k_{ij}=I^k_{ji}$. 
In Eq. (\ref{ini}) $\alpha$ runs over all the third generation scalar quarks
and Higgs scalars. The coefficients in the equation depend on the initial values
of Yukawa and gauge couplings and , of course, are determined by the solutions
of RGEs, (12), (15), (20) and (21).  

The IFQFPs exist for any values of tan$\beta$ if the initial values of all three
Yukawa couplings, $Y^0_k$'s, are large enough (say, larger than the gauge coupling 
at GUT scale), as shown in the last section. However, that whether the condition
can be satisfied is in fact dependent of tan$\beta$ if we want to make the theory
realistic, i.e., to impose the requirement that the third generation quark masses
are given by experiments. Therefore, for numerical calculations, we choose the
procedures as follows. First, run gauge couplings from $m_Z$ scale up to find 
the scale where gauge couplings unify with the SUSY threshold effects taken at
$M_{SUSY}= 400$ Gev. The values of gauge couplings at $m_Z$  
are taken directly from Ref.~\cite{PDG}. For three Yukawa couplings, we
take corresponding current masses of the third generation quarks from Ref. ~\cite{PDG} and calculate their
pole masses, then run them to $m_Z$ scale to find the Yukawa couplings at
that scale with tan$\beta$ fixed (hence the SUSY corrections to the pole mass
of top are not included in our calculations). The values of Yukawa couplings at GUT scale, $Y^0_k$'s, are found
by running their energy scale value up by using Eqs. (7,8,9) and again the SUSY
threshold are taken into account at $M_{SUSY}$. 
The second step is to run the gauge couplings, Yukawa couplings, gaugino masses,
trilinear soft SUSY breaking terms and $\Sigma_l (l= t, b ~and~ \tau)$
down to the low scale (e.g., the $M_{SUSY}$ scale ) with the aid of the integral 
equations (6-9), (12), (15), (20), (21). The masses of the sfermions of the third
and the Higgs bosons can be got via $\Sigma_l(l=t, b ~and~\tau)$ as shown in
Ref.~\cite{km}. With such a procedure the dependences on Yukawa couplings  of the coefficients in
Eqs. (22,\ref{ini}) are translated into the dependence on tan$\beta$.
We show below an example of numerical results at $t_S=log(M^2_{GUT}/M^2_{SUSY})=
63.28$ for tan$\beta=58$:
\begin{eqnarray}
 A_t(t_S) &=& 0.19118 A_t^0 - 0.04580 A_b^0 + 0.01059 A_\tau^0
   - 0.02728 M_1^0 - 0.20187 M_2^0 - 1.45580 M_3^0 , \nonumber  \\
 A_b(t_S) &=& - 0.04019 A_t^0 + 0.06052 A_b^0 - 0.03687 A_\tau^0
   - 0.00168 M_1^0 - 0.14498 M_2^0 - 1.32411 M_3^0, \nonumber  \\
 A_\tau(t_S) &=& 0.03346 A_t^0 - 0.20839 A_b^0 + 0.28668 A_\tau^0
   - 0.08353 M_1^0 - 0.22527 M_2^0 + 0.47013 M_3^0, \nonumber  \\
\Sigma_t(t_S) &=& -0.04580 (\tilde{m}_{D3}^0)^2 - 0.03520 (m_{H1}^0)^2
  + 0.19118 (m_{H2}^0)^2 + 0.01059 (\tilde{m}_{L3}^0)^2
  + 0.01059 (\tilde{m}_{E3}^0)^2 \nonumber  \\
&& + 0.14538 (\tilde{m}_{Q3}^0)^2 + 0.19118 (\tilde{m}_{U3}^0)^2
   - 0.15170 |A_t^0|^2 + 0.00140 |A_b^0|^2 + 0.00227 |A_\tau^0|^2 \nonumber  \\
&& + 0.02344(A_t^0 A_b^{0*}+ A_t^{0*} A_b^0) - 0.00314 (A_t^0 A_\tau^{0*} + A_t^{0*} A_\tau^0)
   - 0.00071(A_b^0 A_\tau^{0*} + A_b^{0*} A_\tau^0) \nonumber  \\
&& + 0.00841(M_1^0 A_t^{0*} + M_1^{0*} A_t^0) - 0.00149(M_1^0 A_b^{0*} + M_1^{0*} A_b^0)
   + 0.00000(M_1^0 A_\tau^{0*}+ M_1^{0*} A_\tau^0) \nonumber \\
&& + 0.04673(M_2^0 A_t^{0*} + M_2^{0*} A_t^0) - 0.00865(M_2^0 A_b^{0*} + M_2^{0*} A_b^0)
   + 0.00072(M_2^0 A_\tau^{0*} + M_2^{0*} A_\tau^0) \nonumber \\
&& + 0.18528(M_3^0 A_t^{0*} + M_3^{0*} A_t^0) - 0.03551(M_3^0 A_b^{0*} + M_3^{0*} A_b^0)
   + 0.00515(M_3^0 A_\tau^{0*} + M_3^{0*} A_\tau^0) \nonumber \\
&& - 0.00418(M_1^0 M_2^{0*} + M_1^{0*} M_2^0) - 0.01949(M_1^0 M_3^{0*} + M_1^{0*} M_3^0)
   - 0.13369(M_2^0 M_3^{0*} + M_2^{0*} M_3^0) \nonumber \\
&& + 0.03411 |M_1^0|^2 + 0.33227 |M_2^0|^2 + 4.96704 |M_3^0|^2, \nonumber \\
 \Sigma_b(t_S) &=& 0.06052 (\tilde{m}_{D3}^0)^2 + 0.02364 (m_{H1}^0)^2
  - 0.04019 (m_{H2}^0)^2 - 0.03687 (\tilde{m}_{L3}^0)^2
  - 0.03687 (\tilde{m}_{E3}^0)^2 \nonumber \\
&& + 0.02032 (\tilde{m}_{Q3}^0)^2 - 0.04019 (\tilde{m}_{U3}^0)^2
   + 0.00420 |A_t^0|^2 - 0.03908 |A_b^0|^2 - 0.00028 |A_\tau^0|^2 \nonumber \\
&& + 0.00441(A_t^0 A_b^{0*} + A_t^{0*} A_b^0) - 0.00225(A_t^0 A_\tau^{0*} + A_t^{0*} A_\tau^0)
   + 0.00859(A_b^0 A_\tau^{0*} + A_b^{0*} A_\tau^0) \nonumber \\
&& + 0.00017(M_1^0 A_t^{0*} + M_1^{0*} A_t^0) - 0.00028(M_1^0 A_b^{0*} + M_1^{0*} A_b^0)
   + 0.00089(M_1^0 A_\tau^{0*} + M_1^{0*} A_\tau^0) \nonumber \\
&& - 0.00243(M_2^0 A_t^{0*} + M_2^{0*} A_t^0) + 0.00974(M_2^0 A_b^{0*} + M_2^{0*} A_b^0)
   - 0.00246(M_2^0 A_\tau^{0*} + M_2^{0*} A_\tau^0) \nonumber \\
&& - 0.01353(M_3^0 A_t^{0*} + M_3^{0*} A_t^0) + 0.05688(M_3^0 A_b^{0*} + M_3^{0*} A_b^0)
   - 0.03167(M_3^0 A_\tau^{0*} + M_3^{0*} A_\tau^0) \nonumber \\
&& - 0.00146(M_1^0 M_2^{0*} + M_1^{0*} M_2^0) - 0.00279(M_1^0 M_3^{0*} + M_1^{0*} M_3^0)
   - 0.09627(M_2^0 M_3^{0*} + M_2^{0*} M_3^0) \nonumber \\
&& - 0.00098 |M_1^0|^2 + 0.23446 |M_2^0|^2 + 4.64976 |M_3^0|^2, \nonumber \\
 \Sigma_\tau(t_S) &=& -0.20839 (\tilde{m}_{D3}^0)^2 + 0.07828 (m_{H1}^0)^2
  + 0.03346 (m_{H2}^0)^2 + 0.28668 (\tilde{m}_{L3}^0)^2
  + 0.28668 (\tilde{m}_{E3}^0)^2 \nonumber \\
&& - 0.17493 (\tilde{m}_{Q3}^0)^2 + 0.03346 (\tilde{m}_{U3}^0)^2
  + 0.00725 |A_t^0|^2 - 0.01117 |A_b^0|^2 - 0.18803 |A_\tau^0|^2 \nonumber \\
&& + 0.00192(A_t^0 A_b^{0*} + A_t^{0*} A_b^0) - 0.00469(A_t^0 A_\tau^{0*} + A_t^{0*} A_\tau^0)
  + 0.07397(A_b^0 A_\tau^{0*} + A_b^{0*} A_\tau^0) \nonumber \\
&& + 0.00035(M_1^0 A_t^{0*} + M_1^{0*} A_t^0) - 0.00932(M_1^0 A_b^{0*} + M_1^{0*} A_b^0)
   + 0.01609(M_1^0 A_\tau^{0*} + M_1^{0*} A_\tau^0) \nonumber \\
&& - 0.00196(M_2^0 A_t^{0*} + M_2^{0*} A_t^0) - 0.01776(M_2^0 A_b^{0*} + M_2^{0*} A_b^0)
   + 0.03236(M_2^0 A_\tau^{0*} + M_2^{0*} A_\tau^0) \nonumber \\
&& - 0.02048(M_3^0 A_t^{0*} + M_3^{0*} A_t^0) + 0.04148(M_3^0 A_b^{0*} + M_3^{0*} A_b^0)
   - 0.04521(M_3^0 A_\tau^{0*} + M_3^{0*} A_\tau^0) \nonumber \\
&& - 0.00595(M_1^0 M_2^{0*} + M_1^{0*} M_2^0) + 0.00716(M_1^0 M_3^{0*} + M_1^{0*} M_3^0)
   - 0.01505(M_2^0 M_3^{0*} + M_2^{0*} M_3^0) \nonumber \\
&& + 0.10639 |M_1^0|^2 + 0.38267 |M_2^0|^2 - 1.72886 |M_3^0|^2 .
\end{eqnarray}

One can see from Eq. (24) that for $A_t$, $A_b$ and $\Sigma_t$, $\Sigma_b$
there are very large coefficients appearing before $M_3^0$ or $|M_3^0|^2$ 
which arise from the large gauge coupling of SU(3) group in the RGEs, 
Eqs. (4) and (5),
and hence make them inevitably greatly depend on the gluino mass. If we
consider the scenario with the mass spectrum below 1 TeV, we find that
the effects of other high energy boundary values on the low energy
spectrum are negligible as illustrated in Eq. (24) and actually it is gluino
mass which mainly govern the SUSY mass spectrum (note that $m^0_{\alpha}\sim$ or less than $M^0_i$ is required due to
the constraint from cosmology~\cite{cos}) so that non-universality has a little influence on low energy mass spectrum.
This is the screening effects induced by the large Yukawa couplings of top, bottom and tau
for which a detailed discussion in the real parameter case has been given in Ref.~\cite{km}. 
For $A_\tau$ and $\Sigma_\tau$, the screening effects are weak since here $M_3^0$ is not as
important as for $A_t$, $A_b$, $\Sigma_t$ and $\Sigma_b$ and the coefficient of $M_2^0$
can compete with that of $M_3^0$. In particular, for $A_{\tau}$, the coefficient of $A_{\tau}^0$ is the same order as
that of $M_3^0$.
 That is, we are in the vicinity of the IRQFP in the example.
Note that a crucial difference between our discussion and the one in Ref.~\cite{km} 
is that we use the Yukawa couplings at $m_Z$ scale induced from their corresponding
current masses of quarks, while in Ref.~\cite{km} Yukawa couplings at the GUT scale much
larger than the unified gauge couplings are assumed when the IRQFP behavior is
discussed for large tan$\beta$. Another difference is that we take the SUSY
threshold at $M_{SUSY}$ whereas in Ref.~\cite{km} SUSY threshold effects was not considered.
As a consequence we find that the value of tan$\beta$ for which the IFQFP is reached
depend weakly on the choice of $M_{SUSY}$. For example the theory exhibits
IRQFP behavior at tan$\beta \simeq 59 $ for $M_{SUSY}=400$ GeV
and at tan$\beta \simeq 60$ for $M_{SUSY}=800$ GeV.

In Fig. 1(a) we illustrate how the dependences of $A_t$, $A_b$ and $A_\tau$
on their initial values, i.e., the coefficient $c_l^l(t_S)$ ( l=t,b,$\tau$)
in Eq. (22), change with tan$\beta$. One may find in the figure that the 
coefficient $c_t^t(t_S)$ increases rapidly for tan$\beta$
less 5 and decrease slowly afterwards. This is because the top Yukawa coupling,
$y_t$, behaves as $\propto \frac{1}{sin\beta}$ and for somewhat large tan$\beta$
it remains almost unchanged. So, according to the procedure described above, 
when tan$\beta$ increases $Y_t^0$ can not reach the value large enough to
make $A_t$ independent of $A_t^0$. On the other hand, the coefficients $c_b^b$ and $c_{\tau}^{\tau}$
 decrease considerably as tan$\beta$ increases
in the whole range and drop sharply to zero when tan$\beta$ approaches 59,
which is exactly the case as expected since their corresponding Yukawa couplings
behave as $\propto \frac{1}{cos\beta}$. When tan$\beta$ reaches some large 
value ( $\simeq 59$ ), $Y^0_{b,\tau}$ become large enough to make
the dependences of $A_{b,\tau}$ on their initial values, $A^0_{b,\tau}$,
almost disappear, i.e., IRQFP is reached.  In Fig. 1(b-d) we plot the the dependences
of $\Sigma_t(t_S)$, $\Sigma_b(t_S)$ and $\Sigma_\tau(t_S)$ on the
initial values of their corresponding three parameters in the definitions of $\Sigma_t$, 
$\Sigma_b$ and $\Sigma_\tau$, i.e., $f_{\alpha}^k (t_S)$, k=t,b,$\tau$ in 
Eq. (\ref{ini}). One can find the similar behavior of the dependences on initial
conditions when changing tan$\beta$. 
It is clear from  Fig. 1 that except for $A_t$ and $\Sigma_t$, IRQFP
behavior becomes more and more evident as tan$\beta$ approaches 59, i.e., 
the point where Landau pole appear as we run bottom and tau Yukawa couplings up. 

In Fig. 2 we plot $\rho_l = A_l/M_3 (l=t, b~and~\tau)$ versus $\alpha_s$ for 
tan$\beta=58.7$, running from the GUT scale down to $M_{SUSY}$ scale as 
indicated by the axes of the strong coupling $\alpha_s$. One can clearly
see the IRQFP behavior of $\rho_l(l=t, b~and~\tau)$ from the figure.
In summary, the numerical results confirm that despite the CP violating phases 
introduced, the model still exhibit the IRQFP behavior for soft terms $A_l$ and $\Sigma_l$,
as pointed out in section two.

Now we come to a position to address the effects of CP violating phases on sparticle 
spectrum. For the effects of phases of gaugino masses, $M_1^0$ and $M_2^0$,
on the mass spectrum(because of R-symmetry, we take $M_3^0$ real for convenience
in the following discussion), one can see from Eq. (24) that the
only way for the phases of $M_1^0$ and $M_2^0$ to be important is that their 
magnitudes are much larger than $M_3^0$(say, at least one order of magnitude),
which means a very heavy bino or chargino mass and so is not a preferable scenario.
For the phases of trilinear terms $A_t^0$, $A_b^0$ 
and $A_\tau^0$, naively one may think that it is possible for them to play 
important role in the mass spectrum through the interference terms with $M_3^0$
as illustrated in Eq. (24). However, notice that because the coefficients of the $A^0_l M^{0*}_3$ terms
are much smaller than that of $|M_3|^2$ term ( e.g., for $\Sigma_t$, the former are only 1/20 of the
latter ) we have to choose $|A_l^0|(l=t, b ~and ~\tau)$ much larger than
$M_3^0$(say, at least one order of magnitude) in order to make the effects of their phases 
considerable so that one is led to far from IRQFP.  Moreover, since trilinear soft SUSY breaking terms appear 
in the nondiagonal terms
of squark or slepton mass matrices, the large $A_l$ induced by very large $A_l^0$
, together with terms proportional to $\mu~ tan\beta$, may result in so light stau that
stau becomes LSP. Therefore, we can infer 
from the above discussion that the initial values of trilinear soft SUSY
breaking terms cannot be much larger than the magnitudes of $M_i^0$.
Thus we can conclude that in physically interesting regions of parameter space where the notion of IRQFP can be applied, 
CP violating phases of gaugino masses and soft trilinear SUSY breaking terms
are not important for determining the mass spectrum through their roles 
played in RGEs.

\section{Conclusion}
In conclusion, we have shown in this paper another way to find the solutions of 
the RGEs of the soft SUSY breaking terms in the third generation sector of MSSM by directly solving the RGEs themselves.
Compared with the method of the spurion expansion, our approach
proves to be a simple and convenient way especially for models with complex
soft trilinear SUSY breaking terms and gaugino masses. The results in the scenario
with complex trilinear terms and gaugino masses turn out to support the 
notion of IRQFP. It follows that non-universality of trilinear couplings $A_l$ and gaugino masses have no influence on
sparticle spectrum at the IRQFP. We have studied the effects on mass spectrum of CP violating phases of trilinear
terms and the non-universal gaugino masses and find that the effects
on the mass spectrum are limited to be small in the physically interesting region of
parameter space where the notion of IRQFP can be applied and physical requirements such as LSP being neutral particle are imposed.

\section*{\bf Acknowledgements} \nonumber
The work was partly supported by National Natural Science Foundation of China.

Figure Caption \\
Fig. 1(a) lines labelled by 1, 2 and 3 correspond respectively  
to coefficient of $A_t^0$ in $A_t(t_S)$, i.e., $c_t^t(t_S)$ in (22), 
coefficient of $A_b^0$ in $A_b(t_S)$, $c_b^b(t_S)$, and coefficient of
$A_\tau^0$ in $A_\tau(t_S)$, $c_{\tau}^{\tau}(t_S)$, as functions of
tan$\beta$. In 1(b) lines 1 and 2 refer respectively to coefficients of
$(\tilde{m}_{Q3}^0)^2$ and $(\tilde{m}_{U3}^0)^2$ in 
$\Sigma_t(t_S)$. In 1(c) lines labeled by 1, 2 and 3 are respectively coefficients of
$(\tilde{m}_{Q3}^0)^2$, $(m_{H1}^0)^2$ and $(\tilde{m}_{D3}^0)^2$ in $\Sigma_b(t_S)$.
In 1(d) lines 1 and 2 correspond respectively to coefficients of $(m_{H1}^0)^2$ and
$(\tilde{m}_{L3}^0)^2$ in $\Sigma_\tau(t_S)$.

Fig. 2 Three dimensional plots of $\rho_l=A_l/M_3(l=t, b~and~\tau)$ as functions
of $\alpha_s$ for tan$\beta=58.7$. The other parameters are chosen as $M_1^0=M_2^0=500$
 GeV, $M_3^0=400$ GeV and $M_0=800$ GeV. Initial values of $|\rho_l|$'s are
chosen as 1 or 2 and the phases of $\rho_l^0$'s are taken as $i\times \pi/4 (i=0,...,7)$.

\begin{figure}
\begin{minipage}[t]{6.1 in}
     \epsfig{file=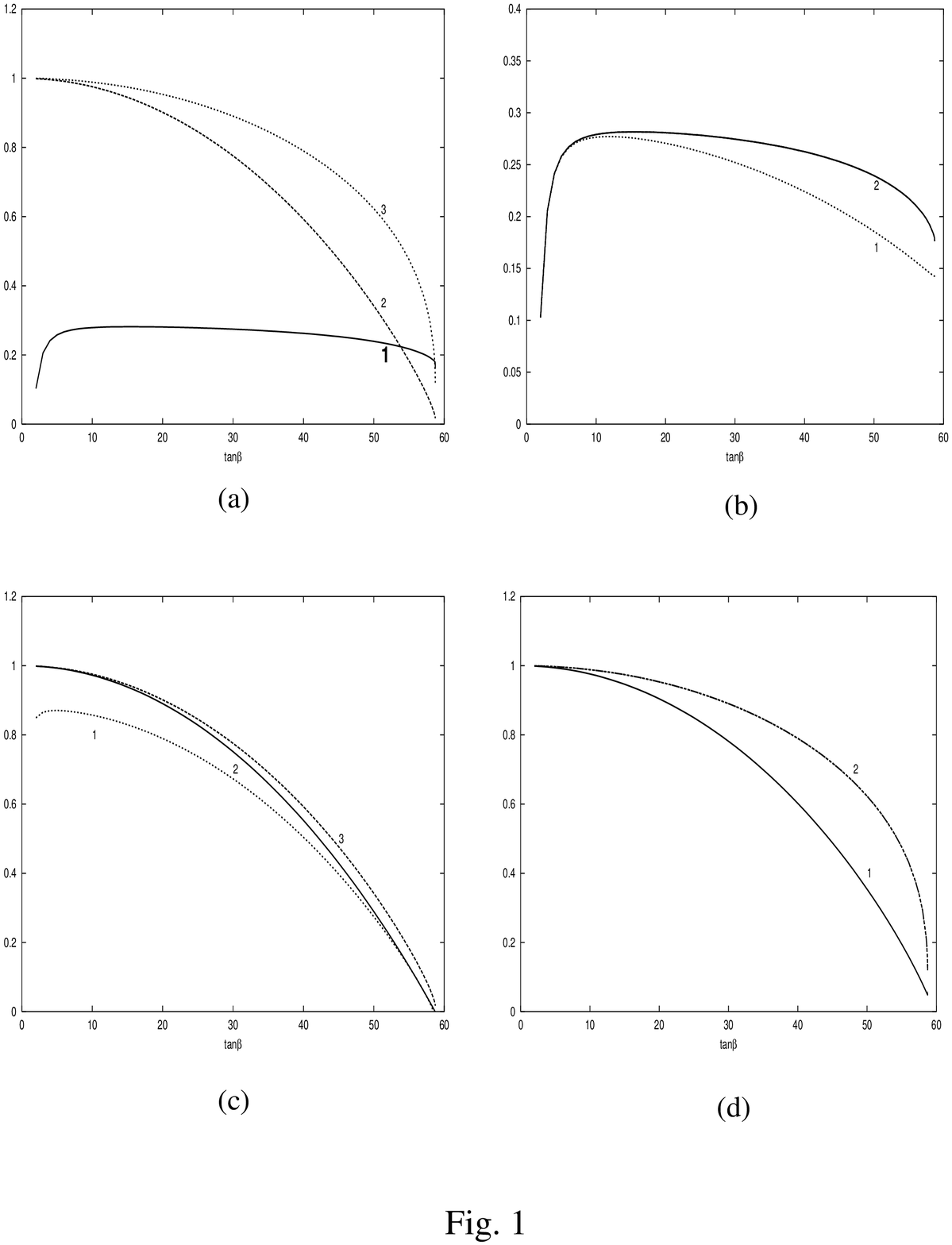,width=7.1 in}
     \end{minipage}
\end{figure}

\begin{figure}
\begin{minipage}[t]{6.1 in}
     \epsfig{file=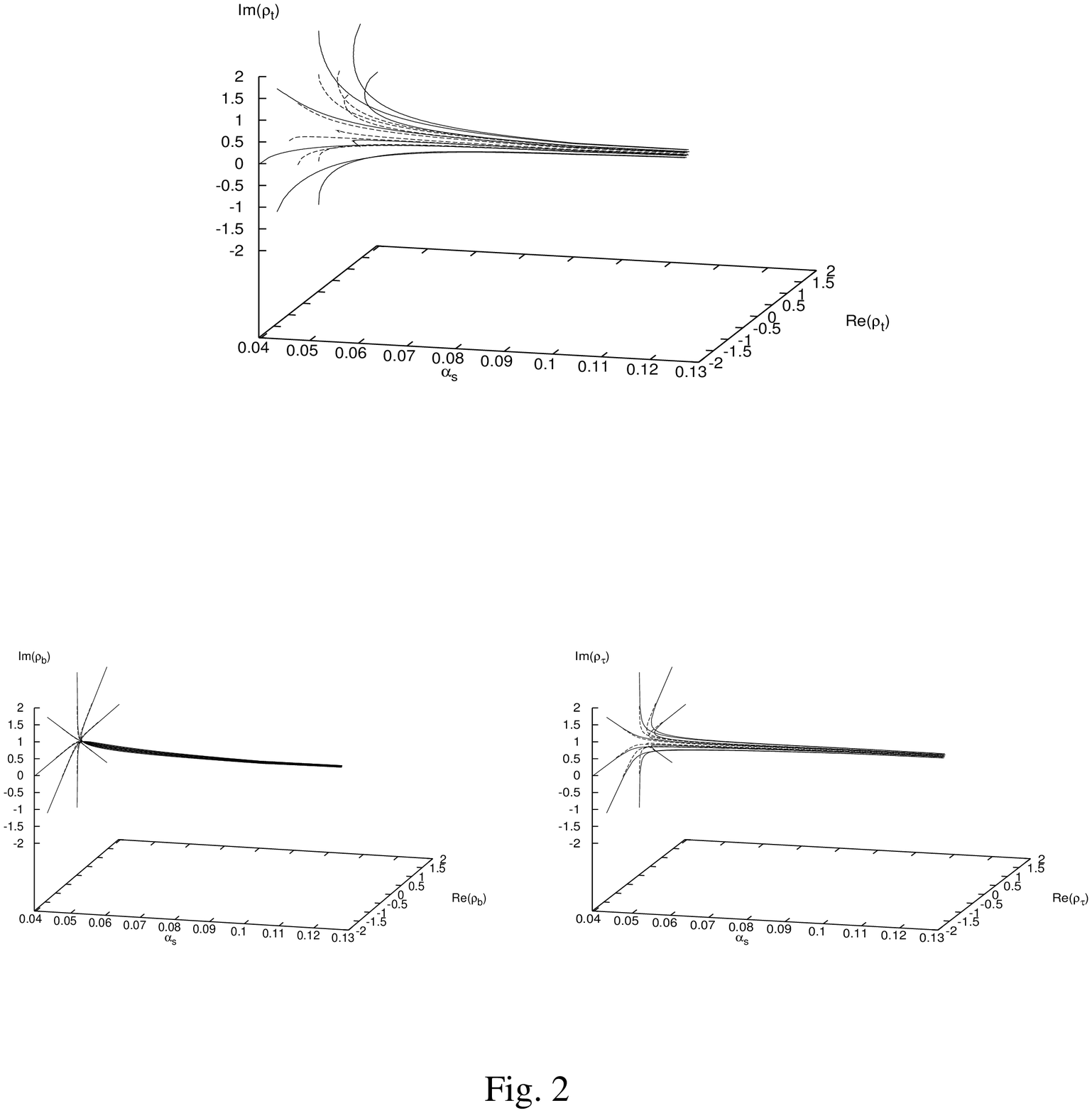,width=7.1 in}
     \end{minipage}
\end{figure}

\end{document}